\documentclass[twocolumn]{revtex4}
\usepackage{graphicx}
\usepackage{amssymb}
\usepackage{tabularx}

\usepackage{amsmath}
\usepackage{xcolor}

\def\De{\Delta}

\def\cL{{\mathcal L}}

\def\cO{{\mathcal O}}

\def\11{{\mathbb 1}}

\def\re{{\rm e}}

\def\rd{{\rm d}}

\def\beq{\begin{equation}}
\def\eeq{\end{equation}}
\def\bea{\begin{eqnarray}}
\def\eea{\end{eqnarray}}

\def\ra{\rightarrow}
\def\Ra{\Rightarrow}

\begin{document}
\title{Searching for supermassive charged gravitinos in underground experiments}
\author{Krzysztof A. Meissner$^1$ and Hermann Nicolai$^2$\\}
\affiliation{\\
$^1$Faculty of Physics,
University of Warsaw\\
Pasteura 5, 02-093 Warsaw, Poland\\
$^2$Max-Planck-Institut f\"ur Gravitationsphysik
(Albert-Einstein-Institut)\\
M\"uhlenberg 1, D-14476 Potsdam, Germany
}

\vspace{10mm}

\begin{abstract}  We examine possible experimental signatures  
that  may be exploited to search for stable supermassive particles with
electric charges of $\cO(1)$ in future underground experiments, and 
the upcoming JUNO experiment in particular. The telltale signal
providing a unique signature of such particles, would be 
a correlated sequence of three or more nuclear recoils along a straight line, 
corresponding to the motion of a non-relativistic ($\beta \lesssim 10^{-2}$) 
particle that could enter the detector from any direction. We provide 
some preliminary estimates for the expected event rates.
\end{abstract}
%\pacs{95.35.+d,04.65.+e}
\maketitle

\vspace{5mm}

%%%%%%%%%%%%%%%%%%%%%%%%%%%%%%%
\section{Introduction.} 
Recent work \cite{MeissnerNicolai2019} has raised the unconventional possibility 
that Dark Matter (or DM, for short) could consist at least in part of an extremely dilute 
gas  of supermassive stable gravitinos with charge $q=\pm\frac23$ in units 
of the elementary charge $e$. Perhaps the most unusual feature of this proposal
is that, due to their large mass and very low abundance such DM candidates can carry 
electric charges of $\cO(1)$ and are thus not dark at all, but luminous, and 
can escape detection only by virtue of their rare occurrence. In principle 
this feature makes these new DM candidates much more  amenable to direct detection, 
provided a way can be found to overcome their low abundance.
By contrast, previously considered DM candidates with masses in the 
TeV range can only carry very small charges (``milli-charges"), and are thus
much harder to detect directly via their putative electromagnetic interactions.

In \cite{MeissnerNicolai2019} we scanned through several past and currently
operating WIMP and monopole searches in order to see whether such 
a particle might already have been sighted in a previous experiment.
As it turned out, the most promising candidate for such reconsideration
is the MACRO experiment \cite{MACRO} which was originally set up to 
search for magnetic monopoles, and which was terminated in 2002.
While no unusual events were reported
in the summary paper \cite{MACRO}, a subsequent search for lightly ionizing 
particles revealed the existence of one outlier event that was subsequently 
discarded by the collaboration, because there appeared to be no way of
deciding between two conflicting interpretations \cite{MACRO1}. 
The more unusual of these interpretations corresponds to a {\em slow} particle 
(not a magnetic monopole) moving upwards, hence having traversed
a large part of the Earth, and possessing a fractional charge \cite{MACRO1}.
In \cite{MN0} we re-examined this event in the light of our proposal
and concluded that it, if confirmed, could support the above DM hypothesis.
In spite of these hints it is clear that the issue cannot be conclusively settled 
on the basis of the existing MACRO data concerning this particular event,
also because \cite{MACRO1} does not provide any details
beyond mentioning that the putative particle is `slow', with a charge of close
to $q\sim \pm\frac23$. Hence the need remains for an independent confirmation or 
refutation of the DM hypothesis proposed in \cite{MeissnerNicolai2019}.

Here we follow up on this observation in order to discuss new and 
independent tests in upcoming or planned  future underground experiments, 
with special attention  to the JUNO experiment \cite{JUNO} which will soon 
start operation with the chief aim of exploring properties of neutrinos and
anti-neutrinos. The JUNO detector is by far the largest underground detector 
ever constructed: it is a liquid scinitillator detector containing 20\,000 tons of
organic fluid,  and the number of scintillators is twenty times bigger than in the 
largest instruments built so far. To eliminate the main background consisting of 
cosmic muons the experiment is located deep underground. In addition
there is an outer water shell providing a veto to muons by the Cherenkov 
radiation and shielding the central detector from natural radioactivity.
Because muons and radioactive processes are thus effectively eliminated
as possible background, the JUNO experiment is ideally suited not only to probe 
(anti-)neutrino physics (its main purpose) but also to search for new types of deeply 
penetrating particles, such as supermassive gravitinos. As we will explain, it
is then rather straightforward to discriminate signals of such particles, 
if they exist, against neutrino induced processes, thus offering unambiguous 
signatures of possible new physics.

Let us therefore briefly recall basic features of our proposal, which has its origins in 
Gell-Mann's observation \cite{GM} that, subject to a `spurion shift' of the 
electric charges, the fermion content of the Standard
Model (SM) can be matched with the spin-$\frac12$ states of the $N\!=\!8$ supermultiplet
associated to maximally extended $N\!=\!8$ supergravity (see also \cite{NW}). 
As a consequence the only extra fermions of the theory beside the usual 48 SM
spin-$\frac12$ fermions would be eight supermassive gravitinos.
Under the SU(3)$_c\,\times\,$U(1)$_{em}$ subgroup of the SM gauge group 
these split as
\beq\label{GravCharges}
         \left({\bf 3}\,,\,\frac13\right) \oplus \left(\bar{\bf 3}\,,\,-\frac13\right)
         \oplus \left({\bf 1}\,,\,\frac23\right) \oplus \left({\bf 1}\,,\, -\frac23\right)
\eeq
and thus all carry fractional electric charges; see \cite{MN2,KM0} 
for more detailed arguments and for a derivation of these quantum numbers.
For the reasons explained in \cite{MeissnerNicolai2019} we will here not be 
concerned with the color triplet gravitinos (see, however, \cite{MN3} for 
possible effects in the physics of ultrahigh energy cosmic rays), but concentrate
on the color singlet gravitinos carrying charge $\pm \frac23$,
which are not subject to strong interactions.
Due to their  fractional charges the gravitinos (\ref{GravCharges}) cannot decay into 
SM fermions, and are therefore stable independently of their mass. 
Their stability against decays makes them natural DM candidates 
\cite{MeissnerNicolai2019}.  From the various bounds on the charge $q$ of 
any putative DM particle of mass $m$ derived in \cite{milli1,milli2,NNP}
we can infer that gravitinos with charges (\ref{GravCharges}) can only 
be viable DM candidates if their mass is close to, but still below, the Planck scale. 

As we explained in \cite{MeissnerNicolai2019} the abundance of color singlet 
gravitinos cannot be determined from first principles, since
they were never in thermal equilibrium due to their extremely small
annihilation cross section, but one can plausibly 
assume their abundance in first approximation to be given by the 
average DM density inside galaxies. From the number given in \cite{WdB} 
it then follows that, if
DM were entirely made out of nearly Planck mass particles, this would
amount to $\sim 3\cdot 10^{-13}$ particles per cubic meter within our solar system.
Notwithstanding, a more accurate estimation of flux rates is hampered by
important uncertainties, for instance possible inhomogeneities in the DM 
distribution within galaxies or stellar systems (where the gravitational 
attraction of the central star could lead to a local enhancement 
or depletion of such particles \cite{DK}). The velocity 
distribution is expected to be centered around the values $10^{-4} \lesssim
\beta \lesssim 10^{-2}$. Indeed, from the virial theorem we expect the 
velocity of these particles in the vicinity of the earth to be $\sim 30$ km$\,\cdot\,$s$^{-1}$ 
if they are bound to the Solar System, and $\sim 300$ km$\,\cdot\,$s$^{-1}$ if they are bound to 
our Galaxy. However, there could be considerable deviations from these numbers,
especially if the DM particles originate from some other source
such as galaxy collisions in the universe.

In this note we discuss the possibility of searching for supermassive
color singlet gravitinos in planned neutrino experiments, pointing out distinctive 
features and signatures relevant for an eventual discovery, and focusing on the 
JUNO experiment. In the remainder we will often refer to the hypothetical
supermassive gravitino simply as `the Particle'.

\section{Interactions of an electrically charged supermassive particle}

According to (\ref{GravCharges}) the DM gravitinos do
participate in SM interactions, unlike for scenarios 
involving supermassive DM particles which are assumed to 
have only weak and gravitational interactions with SM matter.
A supermassive electrically charged particle can easily pass through 
the earth along a straight track and without attenuation or deflection:
the kinetic energy of a (near) Planck mass gravitino with velocity 
$\beta\sim 10^{-3}$ is about $10^{13}$ GeV, so assuming a stopping power 
of $\lesssim 100$ MeV/cm it will lose only a tiny fraction of its energy during
its passage through the earth. In fact, the actual energy loss might be even less in view
of the different interaction properties with ordinary matter, see below.
Under these assumptions, any particle of mass greater than
$10^{14}$ GeV will be able to penetrate the earth. As we assume the 
Particle's mass to be well above this lower bound, all processes
considered here are effectively independent of the Particle's mass.
Hence, to search for such particles and to get rid of unwanted
background, the deeper the detector is located 
underground, the better prospects are for discovery.

Due to its electric charge the Particle would thus interact 
electromagnetically with both electrons and nuclei.
Because of its non-relativistic motion, and neglecting 
spin effects, we are effectively dealing with a point particle of quasi-infinite 
mass, and could thus perform the analysis in the gravitino's rest system.
Therefore the required analysis differs
from standard treatments in the literature
which mostly concern {\em relativistic} particles in interaction with atoms;
the interactions of ordinary `light' (in comparison with Planck 
mass particles) objects and their ionization properties are very well 
understood \cite{PDG}. By contrast,  
what is required here is a proper quantum mechanical 
treatment of a system comprising the Particle (whose mass
can be taken to be infinite for all practical purposes) in interaction 
with a large organic molecule; in this sense, 
we are dealing with a problem in quantum chemistry rather than
relativistic particle physics!  As such 
a calculation is currently not available (but work in progress) let us 
describe the situation in more qualitative terms. Limiting ourselves to
just one atom in the molecule, what 
would we expect this interaction to look like?
There are basically three different regimes as the Particle
passes through an atom. The first possibility is that the Particle
simply `grazes' the atom and interacts only with the outer electron(s).
In the second regime the Particle passes closer to the nucleus, but still
at a distance where the electric charge of the nucleus is partially
shielded by the inner electrons. Finally, in the third regime the Particle 
passes inside the orbit of the closest electron, and in this regime the
interaction effectively takes place between the Particle 
and the nucleus through nuclear recoils (most importantly with
a carbon atom of mass $m_C \sim 12$ GeV/$c^2$).  Here it is 
important to note that due to its large mass, the Particle `can go wherever 
it wants to go' within the atom, without being inhibited in any way by either 
the nucleus or the electron cloud. The main issue in all cases is whether the 
energy released in these processes via photons would be 
enough to excite the photomultiplier tubes (PMTs) of the detector.

For passage sufficiently far away from the nucleus the interaction is mainly 
between the Particle and the outer and more weakly bound electrons. 
A simple classical argument \cite{MN0} shows that, for velocities 
$\beta \lesssim 10^{-3}$ the energy imparted to individual electrons 
by the Particle is much less than the minimum ionization energy, hence no, 
or not much, ionization is expected to occur, in agreement with the lower
part of the Bethe-Bloch curve (Lindhard-Scharff regime) \cite{PDG}.
However, for a positively charged supermassive particle there is another 
possibility: due to the Particle's large mass and its slow motion 
it can spend enough time in the 
electron cloud to `drag along' an electron, forming a fractionally charged 
lightly bound state that moves along undisturbed and with unchanged 
speed (due to the fractional charge the binding energy would be less
than for a hydrogen atom).
Having removed the electron from the atom, it thus leaves behind an ion 
that can be detected by a streamer tube or a drift chamber.
This process can be repeated any number of times along the track because 
the lightly bound electron can be easily stripped off the Particle again.

Closer to the center of the atom, but still sufficiently far away from the 
nucleus the Particle can induce electronic transitions in the inner shells
and lift an electron to a higher excited state. This may give rise to
some fluorescent light, but the light produced in the transition when 
the electron returns to its former state could be too faint to be 
observed by the PMTs. In addition, possible time delays in the 
fluorescent transitions could distort the time identification
along the track.

Finally, if the Particle gets sufficiently close to the nucleus, {\em i.e.}
inside the innermost electron orbit, the collision will look more and more 
like a nuclear recoil, where the Particle hits the nucleus head-on. 
In the Particle's rest frame the process effectively corresponds to (non-relativistic)
Rutherford-like scattering of a point particle (the nucleus) against a fixed point-like
target of infinite mass and charge $q =\pm \frac23$ \footnote{We note 
 that for the strongly interacting color triplet gravitinos in (\ref{GravCharges}) 
 this process could be more disruptive since the Particle could just `smash'  
 the nucleus, which is not possible with only electromagnetic interactions.}.
This is the process that in principle should release enough 
light to be be visible to the PMTs, and this is 
therefore the main kind of event that could be detected by JUNO. 
More precisely, if the Particle's velocity is $\beta$ the maximal velocity that can be
imparted to the nucleus in the recoil is $2\beta$. If the minimal energy of 
the recoil that can be observed is $\sim$ 1 MeV and the recoiling object is a 
carbon nucleus we thus get the minimal $\beta$ for the Particle to be observable
\beq\label{MeV}
2 m_C\beta^2 \, >\,  1\ {\rm MeV}/c^2 \;\Ra \; \beta_{\min}\sim 6\cdot 10^{-3}
\eeq
Recoil from protons would give 12 times lower recoil energy (while the recoil from 
electrons is negligible). The bound (\ref{MeV}) is marginally compatible with
our assumptions on the velocity distribution of supermassive gravitinos
(in the sense that the velocity would have to be on the tail of the distribution).
In addition, a detailed estimate on the event rates depends on the unknown
abundance of such particles, but in the end this will be a matter for
observation to decide.

In summary we expect three kinds of detectable interactions to take place 
along such a track, namely

\begin{itemize}
\item Ionization
\item Electronic excitation without ionization
\item Nuclear recoil
\end{itemize}
As we explained, not all of the interactions, which can take
place simultaneously, are expected to produce signals visible to 
the JUNO detectors (likewise, elastic scattering of the atom as a 
whole would not be expected to lead to a detectable signal). 
Let us also emphasize that these interactions will take place continually 
along the trajectory of the Particle, but only occasionally, such as in the case 
of a quasi-frontal collision with the nucleus, could there be enough 
light to be registered by the PMTs. In other words, we do not expect to 
necessarily see a continuous track, but rather only fragments of a track, with
occasional flashes of light along the track. At any rate, what can be 
seen or not depends crucially on the properties of the scintillator,
and on the sensitivity of the JUNO PMTs and the spectral range accessible
to them (on neither of which we currently have sufficient information).

While the qualitative picture is thus clear, we emphasize again that 
a precise calculation of the process still needs to be done. In
the absence of such a detailed calculation the electric charge of the 
Particle cannot be deduced reliably. Furthermore, it remains an open question 
whether there is any way to confirm that the particle carries spin $s=\frac32$, 
as predicted by (\ref{GravCharges}), via the gravitino's spin dependent
coupling $\cL_{\rm int} \,=\, q \bar\psi_\mu \gamma^{\mu\nu\rho} A_\nu \psi_\rho$.
to the electromagnetic potential.

\section{Prospects for detection}

Past and present neutrino experiments have led to steadily improved
measurements of the (squared) mass differences $\De m^2$ and 
the mixing angles for three neutrino species, although the absolute values
of the neutrino masses remain unknown. These experiments
as well as future planned measurements rely on observations of ionized 
short tracks produced by absorption of (anti)neutrinos by nuclei (nuclear recoil). 
Experiments  such as Kamiokande and SuperKamiokande are based on 
the observation of Cherenkov radiation, while others (SNO, Argon, Xenon) 
are based on the observation of individual flashes produced by the nuclear recoil. 
JUNO will observe both anti-neutrinos from nearby nuclear reactors as well as
neutrinos of astrophysical origin. The former manifest themselves via
the reaction $\bar\nu_e + p \ra n + e^+$ with the subsequent annihilation 
$e^+ + e^- \ra 2\gamma$ and neutron capture, 
while the most relevant reaction in the second case is
$\nu_e + n \ra p + e^-$. Only the latter process produces a signal similar to 
the one expected with the present proposal, but the crucial fact is that
neutrinos will {\em not} produce a {\em sequence} of recoils along
a straight line, so such signals should be easy to discriminate with the 
JUNO detector. Just like for other underground experiments, the 
main background consists of cosmic relativistic muons and the natural 
radioactivity of surrounding rocks. As we already mentioned,
JUNO can manage to eliminate such background.

Assuming the cross section of the nucleus to be of the order of $10^{-30}$ m$^2$ 
and the density of carbon atoms in the scintillator liquid to be 
$5\cdot 10^{28}$ m$^{-3}$ we can estimate the number of recoils per distance as 
\beq
P \; \sim \; 5\cdot 10^{-2}\ {\rm m}^{-1}. 
\eeq
We stress that the geometric cross section represents a very
conservative estimate, which would be valid only for a short range force.
This number may significantly underestimate the actual rate because 
electromagnetic interactions are long range. Sufficiently close to the nucleus where 
the  nuclear charge is no longer screened by electrons we should more
properly assume a Rutherford-like  cross section rather than a 
pointlike cross section, for instance by extending the geometric cross
section to a ball of the radius of the order of the innermost electron orbit.
On the other hand, the momentum transfer to a nucleus can be
smaller than the estimate (\ref{MeV}).
Nevertheless, already with this minimal assumption we can calculate the 
probability of $N$ recoils over length $L$, which is given by a Poisson 
distribution
\beq
p_N \;=\; \frac{(PL)^N}{N!}\re^{-PL}.
\eeq
For a ball of diameter $R$ one must take into account that the length
of passage decreases away from the center of the JUNO detector, so the  
probability of $N$ recoils is 
\beq
p_N(R)=\frac{1}{N!}\int_0^1\rd x\, 2x\,(2PRx)^N\re^{-2PRx}.
\eeq
The JUNO detector has $R\sim 18\,$m, and therefore our formula gives
the following values for the probability of $N$ recoils for the
passage of a {\em single}  particle through the detector:

\vspace{7mm}
\begin{tabularx}{0.4\textwidth} { 
  | >{\centering\arraybackslash}X 
  | >{\centering\arraybackslash}X 
   | >{\centering\arraybackslash}X 
     | >{\centering\arraybackslash}X 
    | >{\centering\arraybackslash}X 
  | >{\centering\arraybackslash}X | }
\hline
PR & N=0 & N=1 & N=2 & N=3  & N=4\\
\hline
0.5 & 0.528 & 0.321 & 0.114 & 0.029 & 0.006\\
\hline
1.0 & 0.297 & 0.323 & 0.214 & 0.105 & 0.041\\
\hline
1.5 & 0.178 & 0.256 & 0.235 & 0.164 & 0.093\\
\hline
\end{tabularx}

\vspace{4mm}\noindent
{\bf   Table 1: recoil probabilities for the passage of a single 
particle through the JUNO detector.}

\vspace{4mm}

Of course, in order to estimate the expected rate of events we still have to 
fold in the estimated flux of supermassive particles (which, as we said, 
is subject to a number of uncertainties). In our previous 
paper \cite{MeissnerNicolai2019} we have estimated the flux as 
\beq
\Phi\,\lesssim \, 0.03 \, {\rm m}^{-2} {\rm yr}^{-1} {\rm sr}^{-1}
\eeq
As we explained in  \cite{MeissnerNicolai2019}  this estimate
is based on the assumption that supermassive gravitinos make up most of 
the DM in the universe, with a local density of $0.3\cdot 10^6$ GeV$\,\cdot\,$m$^{-3}$,
and move at speeds of $\lesssim$ 300 km$\,\cdot\,$s$^{-1}$.
With an area of $\sim \pi (18 {\rm m})^2 \sim 1000$ m$^2$ we thus arrive at 
a first estimate for the number of gravitinos coming from all directions
\beq
\# \; \mbox{(events per year)} \; \sim \; \cO (100) \;\; ,  
\eeq
again modulo the mentioned uncertainties. Let us also stress
that in currently operating DM search experiments which use liquid xenon, 
the analogous estimate is much lower. For instance, for 
XENONnT \cite{Xenon} or LUX-ZEPLIN \cite{LZ}  the estimate
is only $O(0.1)$ per year due to the much smaller area $\sim 1$ m$^2$. 
In view of the velocity bound (\ref{MeV}) we expect only a fraction of 
these gravitinos to produce a visible signal. At any rate,
the distinctive experimental signature would thus be 
a sequence of at least three nuclear recoils along a straight line, 
which can point in {\em any} direction (in particular, going upwards), 
with equal calculated non-relativistic velocity between the measured
subsequent points of recoil (the energies of the subsequent recoils do not 
have to be equal since the collisions do not have to be central). It is therefore
very important to measure as precisely as possible the positions and times
of the events in the detector. There is also the possibility that
the drift chambers would detect the ionizing track consistent both with the straight 
line of nuclear recoils in the central detector (in which case two nuclear recoils
would suffice) and with the reconstructed time -- that would also be a very clear 
and unmistakable signature of a  very heavy charged particle consistent 
with the DM gravitino proposed in \cite{MeissnerNicolai2019}. For $PR\sim 1$ 
we would expect about one triple recoil and two double recoils per year.
Statistics permitting, one might even contemplate plotting the frequency
of detected directions in solar or galactic coordinates, in order to search
for possible correlations.

%%%%%%%%%%%%%%%%%%%%%%%%%%%%%%%%%%%%%%%%%%
\vspace{0.5cm}
\section{Conclusions}
As we have argued, the JUNO experiment offers unique opportunities not just for 
neutrino physics, but also for the possible discovery of new kinds of  particles 
of the type considered here, with the unusual property of having very large mass 
and carrying non-vanishing electric charge of $\cO(1)$.  This is 
because the JUNO detector is located sufficiently deeply underground to
eliminate cosmic ray background other than high energy muons. Furthermore, by 
vetoing these muons, as well as natural radioactivity, it manages to suppress all 
processes other than events induced by neutrinos -- or entirely new types of
particles of the type proposed in this paper! As far as we are aware, 
extending JUNO searches beyond neutrino physics in order to look for supermassive 
gravitinos  would necessitate only relatively minor adjustments of the triggers 
and electronics of the JUNO detector. 

It goes without saying that the discovery of any such event would constitute 
a truly disruptive advance, in the sense that it would be a first clear 
{\em observational} indication pointing towards a novel Planck scale unification 
of matter with gravitation. Experimental vindication of any of the currently
available approaches to quantum gravity and unification has proved
elusive so far, mainly because the Planck scale is so far off that it is difficult 
to imagine {\em any} kind of conclusive test. With the Planck scale out of the reach 
of any conceivable accelerator experiment, underground observatories
such as JUNO may offer a way out of the impasse.

%%%%%%%%%%%%%%%%%%%%%%%%%%%%%%%
\vspace{0.5cm}
\noindent
 {\bf Acknowledgments:} 
 We would like to thank Lothar Oberauer, S{\l}awomir  Wronka and
 Agnieszka Zalewska for helpful doscussions. 
 K.A.~Meissner was partially 
 supported by the Polish National Science Center grant UMO-2020/39/B/ST2/01279.
 The work of  H.~Nicolai has received funding from the European Research 
 Council (ERC) under the  European Union's Horizon 2020 research and 
 innovation programme (grant agreement No 740209).

\end{document}